\documentclass[runningheads]{llncs}

\usepackage[framemethod=TikZ]{mdframed}
\usepackage{paralist}
\usepackage{tikz}
\usepackage{acro}
\usepackage[T1]{fontenc}

\DeclareAcronym{ifc}{
  short = IFC ,
  long = information flow control
}

\DeclareAcronym{cbc}{
  short = CbC ,
  long = correctness-by-construction
}

\usetikzlibrary{positioning}
\title{Quantitative Information Flow Control by Construction for Component-Based Systems}
\titlerunning{Quantitative IFC by Construction for Component-Based Systems}
\begin{document}
\author{Rasmus C. R{\o}nneberg}
\institute{Karlsruhe Institute of Technology, Germany \\
\email{rasmus.ronneberg@kit.edu}}

\maketitle

\begin{abstract}
	Secure software architecture is increasingly important in a data-driven world. 
	When security is neglected sensitive information might leak through unauthorized access.
	To mitigate this software architects needs tools and methods to quantify security risks in complex systems. 
	This paper presents doctoral research in its early stages concerned with creating constructive methods for building secure component-based systems from a quantitative information flow specification.
	This research aim at developing a method that allows software architects to develop secure systems from a repository of secure components.
	Planned contributions are refinement rules for secure development of components from a specification and well-formedness rules for secure composition of said components.
\end{abstract}
      
\section{Research problem}
 
Systems today operate on large amounts of data. This creates value for society, both in terms of scientific and commercial value.
However, we must ensure the \emph{confidentiality} of that data, such that systems do not leak sensitive information through unauthorized access.
And we must consider the \emph{integrity} of systems to ensure that they function correctly and are not influenced by untrusted actors. 
These challenges are especially prominent in security-critical software that handles sensitive information about individuals, such as location data in the mobility domain.
A key challenge is to provide software architects with methods that allow them to build these security-critical systems and reason about security in large and complex systems on an architectural level. 
One framework for analyzing security is \ac{ifc} using a property such as non-interference~\cite{mantel_information_2011}.
Non-interference can be used to analyze whether a specified security policy has been breached, and provide an yes/no answer to this question. 
For instance information flow control has been successfully used to statically check security breaches in programs through types systems~\cite{sabelfeld_language-based_2003,volpano_sound_1996}.  
Non-interference has also been used as the formalism for security in component-based software architectures for cyber-physical systems~\cite{gerking_component-based_2019}.
These types of work enable information flow control at the component and architectural level in a post-hoc analysis. 
We propose building secure systems in an incremental way analogous to the \ac{cbc}~\cite{kourie_correctness-by-construction_2012} approach for functional correctness.
CbC has been raised to the architectural level by enabling correctness-by-construction for component-based system in ArchiCorC~\cite{margaria_scaling_2020}.
However, ArchiCorC only considers functional correctness and not security and information flow control.
This work aim at closing the gap between CbC and information flow control at the architectural level.
An additional challenge is that the previously mentioned work does not support quantification of information flow.
Approaches such as quantitative information flow~\cite{alvim_probabilistic_2010} aim at quantifying exactly how secure or insecure systems are.
Quantification of information flow is generally used in scenarios were some amount of information leakage in a program is allowed, and standard information flow would be too restrictive. 
Currently, CbC approaches can not be used to build systems with quantitative security guarantees.
In this work we want to build software components that are secure by construction, guided by a quantitative information flow security specification, because this yields strong guarantees for the specification in the resulting system, and we want to study which security guarantees we can establish when these components are composed in different ways.
With these results, we want to build and scale a unified way of constructing secure software from the architectural level down to the source code.
In summary the research questions are as follows:
 \begin{mdframed}[roundcorner=5pt]
	\textbf{Main research question:}
	How can we define an incremental approach to constructing secure systems from secure components that adhere to a quantitative information flow policy and ensure security by construction?
\end{mdframed}
This leads to the following subquestions guiding the research methodology.
\begin{mdframed}[roundcorner=5pt]	
	\textbf{RQ1:} How can we devise a constructive approach to building secure software components complying to a quantitative information flow policy?\\
        \textbf{RQ2:} How can we compose these components such that the overall system  adheres to the quantitative security specification?\\
        \textbf{RQ3:} How can we use such an approach to building secure components with different execution and computational models? 
\end{mdframed}

\section{State of the art}
The work related to the above research questions ranges from component-based secure architectures, to the construction of secure components, and finally to quantification of security.
There exist several methods for constructing and analyzing secure component-based architectures using different computational models such as timed automata and labeled transitions systems~\cite{gerking_component-based_2019,mohammad_formal_2011,greiner_non-interference_2016,jasser_back_2018}.
These methods are not constructive and they use non-interference as the formalism of security. In contrast, our approach is constructive and uses quantitative information flow as the formalism for security.

Approaches to building secure software components include post-hoc analysis using both static and dynamic methods and also CbC approaches.
In most post-hoc methods security is ensured through a security type system or a monitor at runtime~\cite{sabelfeld_language-based_2003,volpano_sound_1996,asplund_secure_2021}.
CbC approaches such as in the work by Runge et al.~\cite{runge_immutability_2023,runge_lattice-based_2020} aim at constructing secure programs by refinements of specifications.
However, these approaches do not scale to the architectural level, and they do not study the composition of components. Furthermore, they do not consider quantification of information flow.

Quantitative information flow has been applied to analyse information leakage in programs and as a hyperproperty on the execution traces of programs~\cite{yasuoka_quantitative_2014,sabelfeld_probabilistic_2000}.
But crucially, these works do not consider quantitative information flow as a security specification and do not derive programs through refinements, as this work proposes.
Furthermore, they also do not study the compositional capabilities of quantitative information flow.

\section{Quantitative information flow control by construction }

\begin{figure}[h!]
    \centering
    \includegraphics[width=\textwidth]{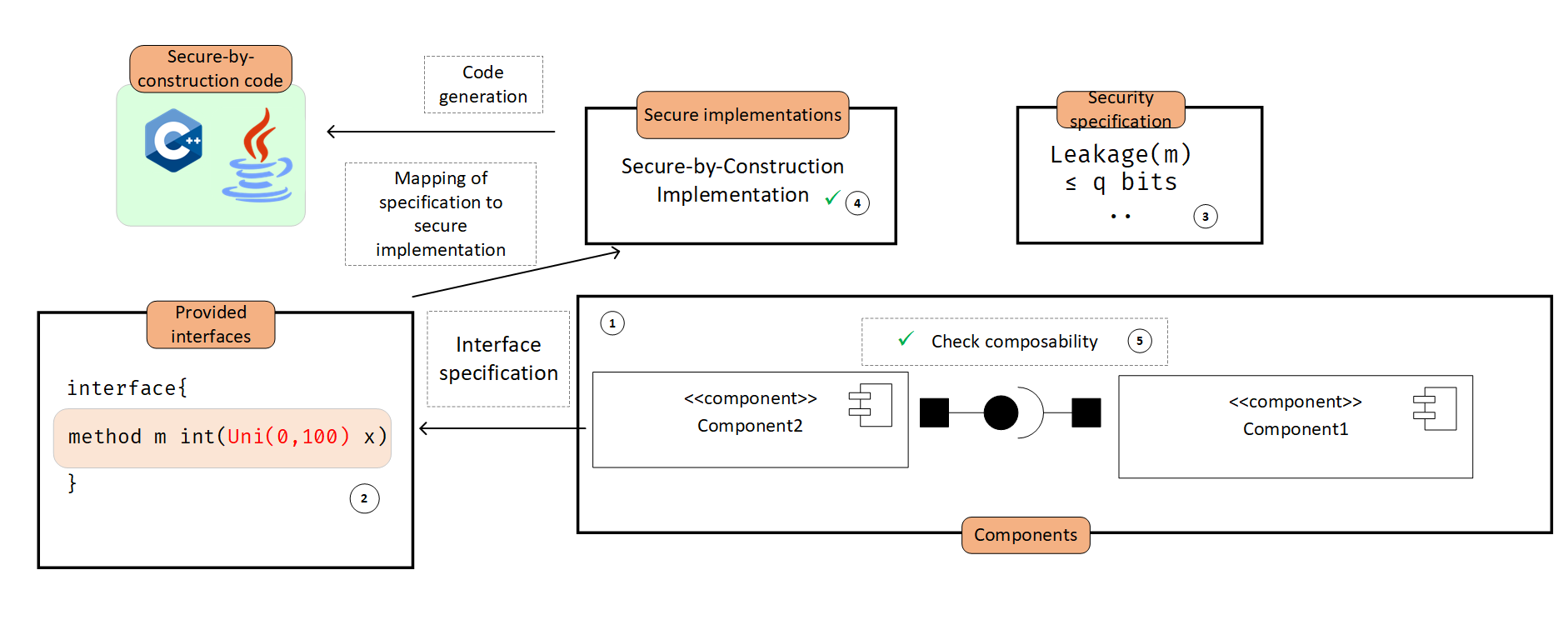}
    \caption{Quantitative Information Flow Control Development Process}
    \label{fig:solution}
\end{figure}

The overarching idea of the information flow control approach is to allow software architects to perform a high-level component-based design of the system.
Components are functionally connected through provided and required interfaces.
Each component is secure-by-construction. The collection of secure components forms a repository that can be used and reused in different settings. 
An overview of the envisioned development process is shown in Figure \ref{fig:solution}.
As shown in step \textcircled{1} we envision that components are connected with other components through required and provided interfaces.
We assume a component model with composite and basic components.
Composite components can contain subcomponents.
Basic components are mapped to the provided interface of said component.
\textcircled{2} A provided interface is a set of method signatures.
The signatures specify a name, return type, and parameters of the method.
Furthermore, they will also contain a specification of the distributions that parameters are assumed to be sampled from, which is necessary for the quantitative information flow control.
\textcircled{3} All methods in the  provided interfaces have a security specification.
The security specification defines an upper bound on how much information can leak from the input to the output of the method.
The upper bounds are specified in terms of quantitative information flow metrics e.g. \emph{Shannon-entropy}, \emph{guessing-entropy}, and \emph{min-entropy}~\cite{alvim_probabilistic_2010}.
\textcircled{4} The methods in the provided interfaces are mapped to secure implementation.
The implementations are constructed incrementally through sound refinement rules.
For each mapping we check that the implementation adheres to the security specification.
\textcircled{5} When components are composed, we check that the composition does not violate any of the security specifications and that the overall system is still secure. 
Finally our approach allows to generate code that is secure by construction.

\section{Research methodology and evaluation}
To realize the proposed solution, we need a mixture of different research methods.
The first step is a \emph{systematic literature review} of state of the art for secure software architectures. 
The second step is to develop the \emph{formal theory} for deriving and composing components from a quantitative information flow policy.
The third step is to build a prototype and to evaluate it using a \emph{case study}.
The forth step is to do an \emph{empirical evaluation} to understand whether the proposed solution can help experts in the field with developing security-critical systems.  

We plan to evaluate our approach on real-life case studies in the mobility domain, where we intend to use our tool to develop software components for an autonomous shuttle operating in a ridepooling setting.
Furthermore, we plan to do expert interviews with software architects and security experts to gather feedback from people in industry.
The expected results of this work are:
\begin{inparaenum}[(1)]
 \item A set of refinement rules for incremental and secure development of components according to a quantitative security specification.  
 \item Well-formedness rules for secure composition of components.
 \item Tool support for developing security-critical software based on the developed concepts.
\end{inparaenum}

Possible limitation of this work that we foresee are a trade-off between usefulness and soundness.
In some cases, we might need to chose between expressiveness and the ability to prove security guarantees.
If successful, this work will bring together correctness-by-construction, secure software architecture, and quantitative information flow.

\subsubsection{Acknowledgments}
This work was supported by funding from the topic Engineering Secure Systems of the Helmholtz Association (HGF) and by KASTEL Security Research Labs

\bibliographystyle{splncs04} 
\bibliography{DocSymRefs} 

\end{document}